\documentclass[twocolumn, 11pt]{article}
\usepackage{array, float}
\usepackage[left=18mm, right=18mm, top=26mm, bottom=24mm]{geometry}
\usepackage{newtxtext}
\usepackage{amsmath, amssymb}
\usepackage{mathrsfs}
\usepackage{tabularx}
\usepackage{graphicx}
\usepackage{dcolumn}
\usepackage{setspace}
\usepackage{pgfpages}
\makeatletter
\usepackage{dcolumn}
\usepackage{url}
\usepackage{comment}
\usepackage{lineno}
\usepackage{cite}

\usepackage{abstract}

\usepackage{setspace}
\usepackage[font=small, format=plain, labelfont=bf, up, textfont=normal, up, justification=justified,singlelinecheck=false]{caption}

\usepackage[affil-it]{authblk} 

\let\OLDthebibliography\thebibliography
\renewcommand\thebibliography[1]{
\OLDthebibliography{#1}
\setlength{\parskip}{0pt}
\setlength{\itemsep}{0pt plus 0.3ex}
}

\title{A noisy opinion formation model with two opposing mass media}
\author{Hirofumi Takesue\thanks{Electronic address: \texttt{hir.takesue@gmail.com}}}
\affil{Faculty of Political Science and Economics, Waseda University}
\date{}

\begin{document}

\twocolumn[

\maketitle

\begin{onecolabstract}
Processes of individual attitude formation and their macroscopic consequences have become an intriguing research topic, and agent-based models of opinion formation have been proposed to understand this phenomenon. This study conducted an agent-based simulation and examined the role of mass media in a noisy opinion formation process, where opinion heterogeneity is preserved by a weak intensity of assimilation and errors accompanying opinion modifications. In a computational model, agents conformed to their neighbours' opinions in social networks. In addition, each agent tended to be influenced by one of a two external agents with fixed opinions, that is, mass media that take opposite positions on an opinion spectrum. The simulation results demonstrated that a small probability of interactions with mass media reduces opinion heterogeneity even with extreme mass media position values. However, a large frequency of interactions with mass media increases opinion heterogeneity. Accordingly, intermediate assimilation strength achieves the least heterogeneous opinion distribution. The influence of mass media dampens the effects of network topology. Our simulation implies that mass media can play qualitatively different roles depending on their positions and intensity of influence. 
\\\\
\end{onecolabstract}
]
\saythanks

\section*{Introduction}
Social influence is a strong determinant of opinion formation. People's opinions are influenced by those around them, such as family members, friends and acquaintances \cite{Mutz2002, Lazer2010, Klofstad2013}. Recent studies have considered the possibility of an echo-chamber, and examined if social influence occurs only among like-minded individuals \cite{Bakshy2015, Barbera2015a, Baumann2020}. Research conducted over the last few decades has found that various attributes and behaviours diffuse on social networks \cite{Christakis2009}.

Agent-based models have become an important research tool to understand opinion formation. Agent-based models identify the rules of micro-level interactions and observe emerging macro-level outcomes. Researchers in multiple fields from sociology to physics have utilised agent-based models to comprehend how people form their opinions in societies \cite{Castellano2009, Flache2017}. In addition, researchers have incorporated a network science framework and examined how local interaction structures defined by social networks determine opinion distribution (see Ref. \cite{Meng2018} for a recent extensive investigation). 

Opinions are often represented on a continuous scale. For example, one's political position or ideology can be arranged on a one-dimensional left--right scale. In another example, attitudes towards specific policies can also be represented by a continuous scale. As the strength of support for a policy varies from person to person, it may require more than a dichotomous variable, for and against, to represent policy attitudes. Previous studies have proposed a continuous opinion model, such as the bounded confidence model \cite{Deffuant2000, Hegselmann2002}. 

Assimilative interaction is one tendency that has been incorporated in various opinion formation models \cite{Axelrod1997, Deffuant2000, Hegselmann2002}. People tend to become more agreeable when they are surrounded by those who have different opinions. Opinion formation models have found that agents' opinions tend to converge after interactions. This basic pattern can be understood as people minimising the dissonance created when holding different opinions \cite{Groeber2014}. 

Opinion distributions demonstrate diversity. The typical character of opinion distribution is a central peak at the middle position exemplified by people's ideological positions on a left--right scale \cite{Flache2017}. At the same time, an existing central peak does not indicate opinion homogeneity; opinions are distributed around the centre and remain diverse. This pattern does not emerge in canonical models of continuous opinion formation. In the bounded confidence model, for example, opinions converge into several clusters, and opinions of all agents can be represented by one of those clusters \cite{Deffuant2000, Hegselmann2002}. In addition, a small probability of interacting with agents outside those clusters leads to a totally homogeneous state \cite{Mas2010}. A loss of diversity due to interaction noise is also observed in discrete models of social influence \cite{Klemm2003a, Macy2015}, and a different form of social influence \cite{Flache2011a} or a learning rule \cite{Banisch2019} is required to maintain diversity.

One method to achieve opinion diversity is to introduce errors in opinion modifications. Added noise preserves opinion heterogeneity by working against the tendency to assimilation. Some studies have confirmed that errors in opinion modifications preserve the original patterns of bounded confidence models; overly large errors dissolve existing clusters and induce a transition to disordered states \cite{Pineda2009, Grauwin2012, Kurahashi-Nakamura2016}. In these models, however, clusters generated by restricted interaction mainly preserve opinion heterogeneity. Another model \cite{Mas2010} assumes that noise becomes larger when interactions occur between similar agents. The motivation to differentiate from similar individuals may be more appropriate for modelling fads such as fashion than for modelling opinion formation. 

In this study, we examined the determinants of opinion distribution when noisiness preserved opinion heterogeneity. We expanded a model introduced in Ref. \cite{Laver2020a}. This model assumes assimilative interaction in the same way as many models. However, a small intensity of assimilation is assumed, and the speed of opinion convergence becomes smaller. In addition, fixed-size errors accompany the opinion modification process and prevent the perfect convergence of opinion. The combination of weak assimilation intensity and errors preserves opinion heterogeneity. We investigated whether opinion polarisation depends on system parameters when some extent of opinion heterogeneity is always preserved by a noisy process (please refer to \cite{Schweighofer2020a}; however, for a different aspect of polarisation, we focused on opinion heterogeneity). 

Another influential factor we considered in our opinion formation model was mass media. In opinion formation models, mass media are treated as external agents who interact with many agents \cite{Shibanai2001}. Mass media tend to preserve opinion diversity, despite interacting with many agents globally \cite{Shibanai2001, Gonzalez-Avella2005, Gonzalez-Avella2010, Peres2010, Peres2011, Pineda2015, Cosenza2020}. One branch point in modelling is the number of mass media. Although earlier studies considered single mass media, recent studies have considered the effects of multiple mass media \cite{McKeown2006, Quattrociocchi2014, Bhat2019}. Competition between multiple mass media is a basic component of advanced democracies; therefore, we examined how two mass media outlets with opposite opinions influenced opinion heterogeneity. This setting has some similarity with models that assume extremists at the edge of an opinion spectrum \cite{Deffuant2006, Mathias2016}.

In the following section, we introduce our opinion formation model. The results section first examines the model's basic behaviour without mass media because Ref. \cite{Laver2020a} examined complex cases where multiple groups existed in the population. We examined the basic behaviour of a simpler model where agents were not distinguished by group membership. After that, we introduced mass media and examined their effects on opinion heterogeneity. The last section summarises the results and discusses the potential extensions of this study.

\section*{Model}
We considered a computational model where $N$ agents were located on a small-world network \cite{Watts1998}. When generating networks, we first considered an expanded cycle where agents were arranged in a circle and were connected with $d / 2$ neighbours on both sides; then, we added $Ndp / 2$ random links \cite{Newman2000}. Loops and duplicated links were avoided. Following this procedure, the average degree size, that is, the number of neighbours, was $d (1 + p)$. A larger $p$ indicated that the network had become more disordered. Previous studies have reported that disordered networks tend to induce consensus in different opinion formation models (see Ref. \cite{Flache2011} and cited studies). 

Each agent had an attribute called an opinion. The opinion of agent $i$ was denoted as $o_i \in \mathbb{R}$. Opinions could represent attitudes towards a specific policy. Positive and negative opinion values represented those for and against that policy; absolute opinion values indicated attitude strength. In the initial states, $o_i$ followed a uniform distribution, $\rm{U} (-3, 3)$. Agents could update their opinion through interactions with other agents. During each elementary time step, one agent could modify his/her opinion; in other words, we assumed asynchronous updating. During each step, an interaction with a neighbour occurred at a probability of $1-w$, whereas interaction with mass media occurred at a probability of $w$. Therefore, a larger $w$ indicated a stronger influence of mass media.

Agents modified their opinion by interacting with their neighbours. During such an event, one \textit{focal} agent, $i$, was selected randomly from $N$ agents; one \textit{role} agent, $j$, was also selected randomly from $i$'s neighbours. We assumed an assimilative interaction, and that $i$'s opinion was influenced by $j$'s opinion. Specifically, the opinion of the focal agent was updated in the following way:
\begin{equation}
o_i \leftarrow o_i + \mu (o_j - o_i) + e,
\end{equation}
where $\mu$ ($0 < \mu \leq 1$) was the intensity of assimilation. We basically assumed small $\mu$ values, which indicated the limited effects of persuasion. In addition to assimilative interaction, random errors denoted by $e$ also modified opinions. A random error term, $e$ took the value $\Delta$ or $-\Delta$ with equal probability. This error term allowed for the possibility that the focal agent's opinion diverged from the role agent's opinion, which reflected the uncertain effects of persuasion. Although the original model \cite{Laver2020a} assumed that assimilation occurred after adding an error term, we just added two components to permit a separate analysis of assimilation and errors. 

In this model, mass media were external agents who influenced the opinions of other agents in networks, but did \textit{not} modify their own opinions. In this model, we assumed that there were \textit{two} mass media with competing opinions \cite{McKeown2006}. For instance, two mass media outlets may represent left- and right-wing opinions. Therefore, we assumed that the opinions of two mass media outlets were $o_m$ and $-o_m$, respectively. The fixed positions of mass media implied that mass media opinions changed much more slowly.

Agents in networks modify their opinions through interactions with mass media as well as with neighbours. During this event, one focal agent was selected similarly to when interactions with a neighbour occurred; then, mass media functioned as a role agent. Mass media each agent followed were fixed during a simulation run. This setting meant that agents kept following left-oriented mass media (e.g. MSNBC) or right-oriented mass media (e.g. FOX). At the beginning of a simulation run, each agent was assigned to either mass media outlet with equal probability. When interactions occurred, agents updated their opinion using the same method as during interactions with their neighbour. In other words, agent $i$ updated his/her opinion in the following way:
\begin{equation}
o_i \leftarrow o_i + \mu (o_m^{(i)} - o_i) + e,
\end{equation}
where $o_m^{(i)}$ was an opinion of mass media agent $i$ followed, and it took $o_m$ or $-o_m$.

Our main interest was the standard deviation of opinions that was denoted by $\hat{\sigma}_o$. Smaller values suggested that people approached consensus states, whereas larger values suggested polarised opinions. After a sufficiently long relaxation process that continued for from 160000$N$ to 50000000$N$ rounds, we recorded the quantity of interest for from 30000$N$ to 420000$N$ rounds. To enhance the statistical accuracy of the simulation results, we conducted at least ten simulation runs and reported the mean values of these outcomes.

\section*{Results and Discussion}
\begin{figure}[tbp]
\centering
\vspace{5mm}
\includegraphics[width = 85mm, trim= 20 20 0 0]{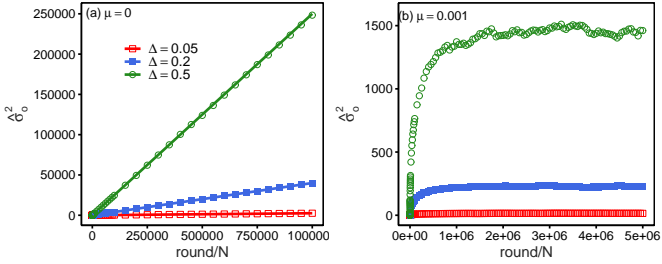}
\caption{\small $\hat{\sigma}_o^2$ is reported as a function of time. Panel (a) presents the results without assimilation ($\mu = 0$). The time evolution of the variance could be approximated by the linear time trend ($\Delta^2t$), where $t$ was the number of rounds divided by $N$. In contrast, panel (b) shows that $\hat{\sigma}_o^2$ reached stable values once assimilation was considered ($\mu = 0.001$). Note that the maximum value of the vertical axis is smaller in panel (b). Parameters: $N = 1000, d = 4$ and $p = 0.01$.}
\label{fig_sigma2_mcs}
\end{figure}

First, we checked the basic model characteristics when interactions with mass media were not considered. Figure~\ref{fig_sigma2_mcs} presents opinion variance ($\hat{\sigma}_o^2$) as a function of time. In this figure, we set the initial opinions of all the agents to 0 to clarify the time trend. The left panel presents the results when no social influence was involved with opinion formation ($\mu = 0$). Here, the model was reduced to a one-dimensional random walk whose movement unit size was $\Delta$. The time evolution of the variance of simulated opinions could be approximated by a linear time trend. In contrast, we assumed weak assimilation ($\mu = 0.001$) in panel (b). Opinion variance reached stable values, and an unlimited increase in variance was prevented once assimilation was introduced (note that the maximum value of the vertical axis is smaller in panel (b)). Therefore, we assumed (weak) assimilation below and investigated the stationary heterogeneity of simulated opinions.

\begin{figure}[tbp]
\centering
\vspace{5mm}
\includegraphics[width = 75mm, trim= 20 20 0 0]{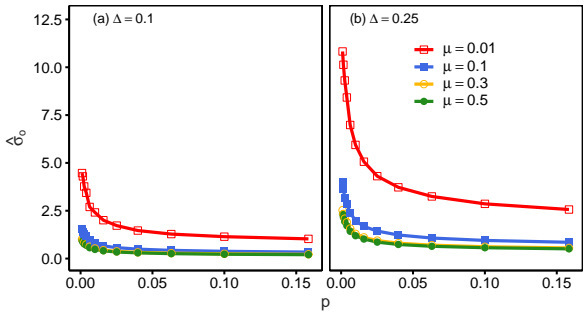}
\caption{\small $\hat{\sigma}_o$ is reported as a function of $p$. A larger $p$ decreases the values of $\hat{\sigma}_o$, which indicates that disordered networks tend to reach a less heterogeneous opinion distribution. Parameters: $N = 1000$ and $d = 4$.}
\label{fig_sigma_p}
\end{figure}

To understand the model's basic behaviour, Figure~\ref{fig_sigma_p} reports the values of $\hat{\sigma}_o$ as a function of a parameter of small-world networks ($p$). Small increases in $p$ significantly lower the values of $\hat{\sigma}_o$. Although further increases in $p$ lead to smaller $\hat{\sigma}_o$, its marginal effects become smaller. Our result was consistent with previous studies that observed that disordered networks tend to move towards consensus \cite{Flache2011}. The effects of other parameters are intuitive; a stronger tendency towards assimilation ($\mu$) reduces opinion heterogeneity, whereas a larger error ($\Delta$) increases opinion heterogeneity.

\begin{figure}[tbp]
\centering
\vspace{5mm}
\includegraphics[width = 85mm, trim= 20 20 0 0]{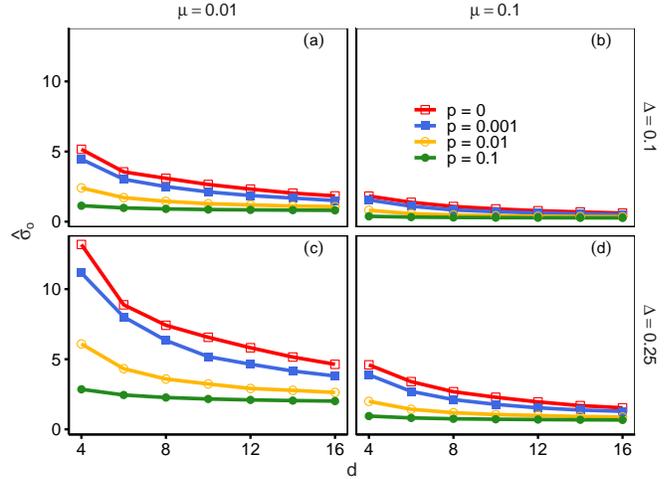}
\caption{\small $\hat{\sigma}_o$ is reported as a function of $d$ for different values of $p$. A larger $d$ decreased the opinion variance. However, small increases in the values of $p$ had the same effect size, which suggests that disordered networks are efficient at decreasing opinion heterogeneity. Parameters: $N = 1000$.}
\label{fig_sigma_d}
\end{figure}

Figure~\ref{fig_sigma_d} reports the values of $\hat{\sigma}_o$ as a function of $d$ for different values of $p$. The values of $\hat{\sigma}_o$ decreased as $d$ took larger values, which means that a larger degree size decreased opinion heterogeneity. At the same time, this figure also shows that adding connections to the nearest neighbours had smaller effects than adding \textit{random} links by increasing $p$. For example, panels show that an opinion approached consensus by increasing the values of $d$ from 4 to 14 when $p = 0$. Almost the same magnitude of heterogeneity reduction was achieved by increasing the value of $p$ from 0 to 0.01, but this modification only increased the average degree from 4 to $4(1 + 0.01)$. Therefore, this figure suggests that disordered networks are efficient at achieving a less polarised opinion distribution.

\begin{figure}[tbp]
\centering
\vspace{5mm}
\includegraphics[width = 85mm, trim= 20 20 0 0]{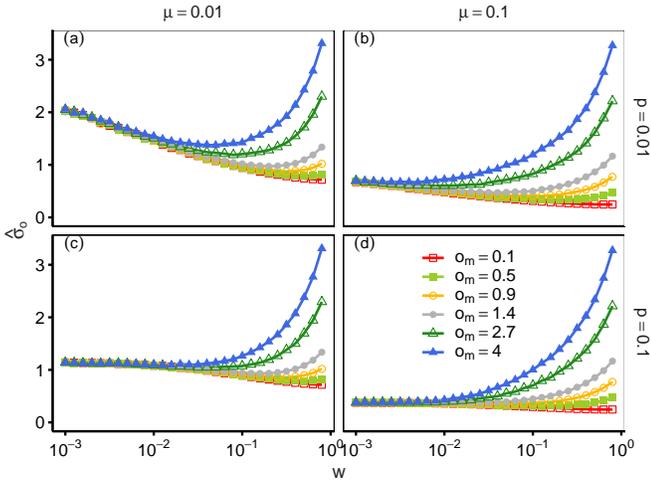}
\caption{\small $\hat{\sigma}_o$ is reported as a function of $w$. Opinion heterogeneity decreased as $w$ took larger values, and the size of this heterogeneity reduction did not depend on the position of mass media ($o_m$). However, $\hat{\sigma}_o$ started to increase when $w$ became overly large, which indicates the non-monotonic impact of the frequency of interacting with mass media. Parameters: $N = 1000, d = 4$, and $\Delta = 0.1$.}
\label{fig_sigma_w}
\end{figure}

From here, we introduced mass media and examined how they contribute to the resulting opinion distribution. Figure~\ref{fig_sigma_w} reports $\hat{\sigma}_o$ as a function of $w$. Introducing opportunities for interactions with mass media reduces opinion heterogeneity. The values of $\hat{\sigma}_o$ decreased as $w$ took larger values, and this pattern was more notable when assimilation was weak (small $\mu$) and networks were less disordered (small $p$). Where $w$ was not overly large, the figure shows that the extremity of media ($o_m$) whose values range from zero to four had a small impact on the extent of heterogeneity reduction. However, further large $w$ increased opinion heterogeneity whose extent depended on the value of $o_m$.

\begin{figure}[tbp]
\centering
\vspace{5mm}
\includegraphics[width = 85mm, trim= 20 20 0 0]{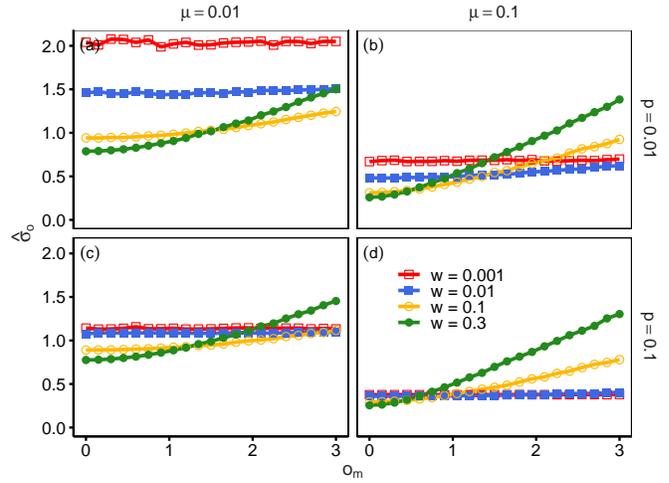}
\caption{\small $\hat{\sigma}_o$ is reported as a function of $o_m$. The values of $o_m$ had a small impact when $w$ was small, which indicated that mass media positions had negligible effects on opinion heterogeneity as long as an interaction probability with mass media was small. Opinions became heterogeneous when $o_m$ and $w$ took large values. Parameters: $N = 1000, d = 4$ and $\Delta = 0.1$.}
\label{fig_sigma_om}
\end{figure}

Figure~\ref{fig_sigma_om} reports $\hat{\sigma}_o$ as a function of $o_m$ to further understand the role of mass media. The figure shows that $o_m$ had a small impact on opinion heterogeneity as long as the value of $w$ was sufficiently small as suggested in Fig.~\ref{fig_sigma_w}. In contrast, $\hat{\sigma}_o$ increased monotonically with the size of $o_m$ when the probability of interaction with mass media was larger. Therefore, a large $w$ led to less opinion heterogeneity when $o_m$ was small, but an opposite pattern was observed with larger values of $o_m$.

\begin{figure}[tbp]
\centering
\vspace{5mm}
\includegraphics[width = 85mm, trim= 20 20 0 0]{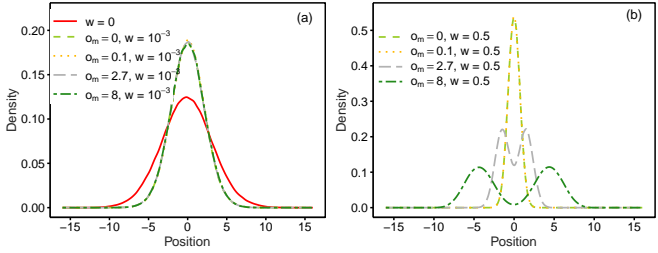}
\caption{\small Distributions of opinions are reported. Panel (a) shows that the introduction of interactions with mass media ($w = 10^{-3}$) reduced opinion heterogeneity. This effect was observed with extreme mass media positions ($o_m = 8$). In contrast, panel (b) shows that a stronger influence of mass media ($w = 0.5$) can induce polarisation. Parameters: $N = 1000, d = 4, p = 0.01, \Delta = 0.1$ and $\mu = 0.01$.}
\label{fig_dist}
\end{figure}

Figure \ref{fig_dist} presents opinion distributions to gain a clearer understanding of the effects of mass media. We conducted 500 simulation runs and recorded opinions at the $500000N$th round to report this result. Panel (a) presents the results with weak mass media influence. Opinion distributions shrank when a small probability of interacting with mass media was introduced ($w = 10^{-3}$). Mass media positions had a negligible impact on the distributions. In this figure, the largest value of $o_m$ was eight. Although this value lies at almost the edge of opinion distributions when $w = 0$, the introduction of mass media still decreased opinion heterogeneity. In contrast, panel (b) shows that opinion distributions polarised when the interaction with mass media was frequent ($w = 0.5$). At this stage, the mass media have an overly high influence, and their positions are reflected in opinion distributions. As the positions of mass media became more extreme, we observed the two peaks at more extreme positions.

These results suggest that the mass media play a different role depending on the values of $w$. The mass media generated indirect connections between agents because half of the agents were influenced by one media and that influence diffused through network interactions. Mass media played a similar role to the additional links in networks (i.e. larger $p$ or $k$) and reduced opinion heterogeneity. At this stage, mass media positions played a relatively minor role as suggested by the limited effects of $o_m$ in Figures~\ref{fig_sigma_w} and \ref{fig_sigma_om}. In contrast, polarised mass media opinions exerted their influence once $w$ became sufficiently large. At this stage, agents' opinion formation was dominated by mass media opinions, and the polarisation of mass media contributed directly to opinion heterogeneity among agents. 

\begin{figure}[tbp]
\centering
\vspace{5mm}
\includegraphics[width = 85mm, trim= 20 20 0 0]{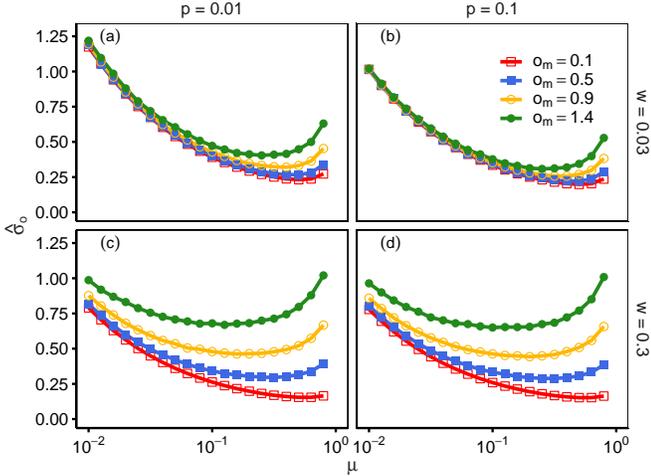}
\caption{\small $\hat{\sigma}_o$ is reported as a function of $\mu$. An increase in $\mu$ decreased opinion heterogeneity as long as $\mu$ values were small. However, a further increase in $\mu$ led to polarised opinions, which indicated the non-monotonic effect of assimilation intensity. Parameters: $N = 1000, d = 4$ and $\Delta = 0.1$.}
\label{fig_sigma_mu}
\end{figure}

Non-monotonic patterns can be observed with another system parameter. Figure~\ref{fig_sigma_mu} presents the effects of assimilation strength ($\mu$). As the values of $\mu$ became larger, a decline in opinion heterogeneity was observed; stronger assimilation made opinion distributions approach consensus states. In contrast, further increases in $\mu$ led to an increase in opinion heterogeneity. This reversal tends to be prominent when mass media positions became extreme, that is, larger $o_m$. External agents with a fixed opinion attracted followers, which reversed the direction of the effects of the assimilation parameter. 

\begin{figure}[tbp]
\centering
\vspace{5mm}
\includegraphics[width = 85mm, trim= 20 20 0 0]{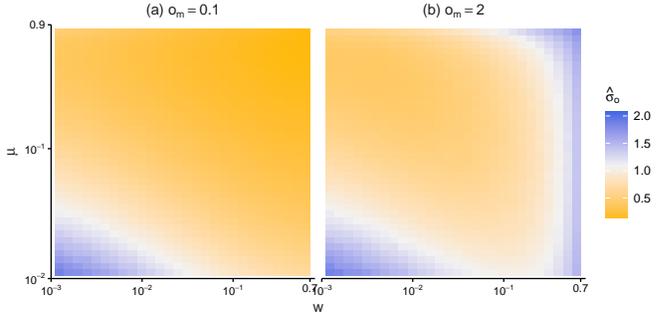}
\caption{\small $\hat{\sigma}_o$ is reported as a function of $w$ and $\mu$. Panel (a) shows that both parameters reduced opinion heterogeneity monotonically when $o_m$ was small. In contrast, panel (b) shows that two parameters demonstrated a non-monotonic impact when $o_m$ was large. Parameters: $N = 1000, d = 4, p = 0.01$ and $\Delta = 0.1$.}
\label{fig_sigma_w_mu}
\end{figure}

Figure~\ref{fig_sigma_w_mu} reports opinion heterogeneity as a function of $w$ and $\mu$ to fully understand the effects of these two parameters. Two parameters demonstrated a monotonic impact when the value of $o_m$ was small. Panel (a) shows that $\hat{\sigma}_o$ decreases as $w$ and $\mu$ increase, which indicates that both the frequency of interaction with mass media and intensity of assimilation contribute to reduced opinion heterogeneity. 

In contrast, non-monotonic patterns appeared when $o_m$ was large. Panel (b) shows that $\hat{\sigma}_o$ had the smallest values with intermediate $w$, and this non-monotonic relationship was prominent with smaller values of $\mu$. When $\mu$ was sufficiently large and $\hat{\sigma}_o$ took very small values, higher frequencies of interaction with mass media monotonically increased opinion heterogeneity. 
 
\begin{figure}[tbp]
\centering
\vspace{5mm}
\includegraphics[width = 85mm, trim= 20 20 0 0]{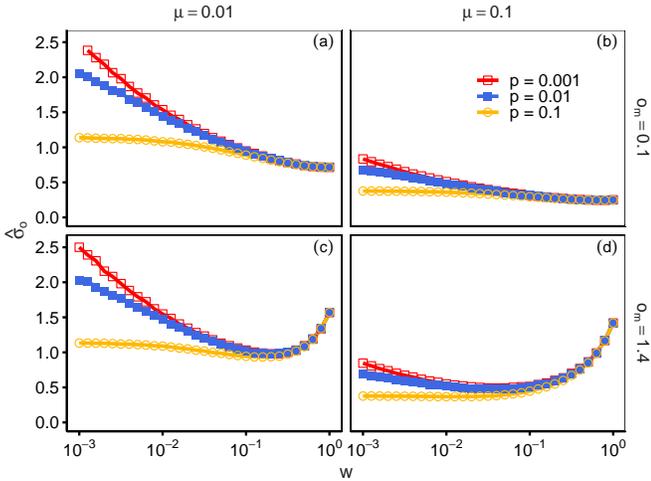}
\caption{\small $\hat{\sigma}_o$ is reported as a function of $w$ for different values of $p$. The dependence of opinion heterogeneity on $p$ diminished as $w$ became larger, which indicated that a larger probability of interaction with mass media suppressed the role of long-range links. Parameters: $N = 1000, d = 4 $ and $\Delta = 0.1$.}
\label{fig_sigma_w_p}
\end{figure}

Finally, we examined how the introduction of mass media modified the dependence on other parameters. Figure \ref{fig_sigma_w_p} reports the values of $\hat{\sigma}_o$ as a function of $w$ for different values of $p$. As reported in Figure \ref{fig_sigma_p}, disordered networks reduced opinion heterogeneity without the influence of mass media. However, this effect diminished as $w$ became larger, and the difference between the results with $p = 0.001$ and $p = 0.1$ almost disappeared when the value of $w$ was approximately 0.1. This figure shows that interactions with mass media can dominate other parameters even when the majority of interactions occur between neighbours. 

\section*{Conclusion}
In sum, we extended and examined the noisy opinion formation model proposed in Ref. \cite{Laver2020a}, where weak assimilation intensity and errors preserved opinion heterogeneity. Our model focused on the role of external agents with fixed opinions, that is, mass media. First, we examined the basic behaviour of the model in the absence of mass media. Opinion heterogeneity converged even with weak assimilation intensity. Additional links, especially long-range links, reduced opinion heterogeneity significantly. Next, we introduced mass media and examined their impact on opinion distributions. The frequency of interaction with mass media indicated non-monotonic effects on opinion heterogeneity. Small increases in interaction frequencies reduced opinion heterogeneity, and this pattern could be observed even with extreme mass media positions. The effects of mass media on opinion positions were negligible with a lower frequency of mass media interaction. At this stage, mass media played the same role as additional network links and contributed to the reduction of opinion heterogeneity. In contrast, more frequent interactions with mass media increased opinion heterogeneity. High frequencies of mass media interaction attracted agents' opinions and contributed to polarisation. The assimilation intensity also demonstrated non-monotonic effects, and both weak and strong intensities contributed to opinion heterogeneity. Finally, higher frequencies of interaction with mass media dampened the effects of network-related parameters. 

Our simulation results imply that multiple mass media with opposite positions can play qualitatively different roles. A moderate influence of mass media can reduce opinion heterogeneity by offering a shared basis for people's opinions, although mass media do not occupy the centre of an opinion spectrum. Conversely, mass media can foster opinion heterogeneity when its influence becomes overly large. Empirical studies have found that mass media influence explains the recent rise of polarisation \cite{Hopkins2014, Martin2017}. Our simulation suggests that the mass media's strong influence attracts followers to extreme positions. 

Finally, our study had some limitations that have implications for possible future studies. We assumed a tendency towards assimilation including in interactions with mass media. However, some studies have also considered the possibility of repulsion in opinion formation, and this modification produces some novel system behaviour \cite{Jager2005, VazMartins2010, Flache2011, Chen2017d, Huet2018, Turner2018}. Although the empirical validity of opinion repulsion remains ambiguous according to recent experimental studies \cite{Takacs2016, Bail2018, Guess2018}, we can examine whether the introduction of repulsion is useful when replicating stylised facts. In addition, network links were fixed in this simulation. Dynamic networks tend to produce opinion diversity \cite{Centola2007, Kozma2008}, but diversity can disappear if errors are introduced in the link updating process \cite{Grauwin2012}. Links with mass media were also fixed in this study, but these links may also be modified considering selective exposure \cite{Stroud2008, Iyengar2009}. Although mass media had negligible effects on opinion heterogeneity when the interaction frequency was low, dynamic relationships with mass media may lead to different patterns. Although this study had these limitations, we believe the basic patterns reported here would be a good basis to investigate more complex situations. 

%\bibliographystyle{unsrt} 
%\bibliography{BIB}

\end{document}